\documentclass{aa}  
\usepackage[varg]{txfonts}
\usepackage{graphicx}
\graphicspath{{Figures/}}
\usepackage{txfonts}
\usepackage{float}
\usepackage{hyperref}
\usepackage{color}
\usepackage[usenames,dvipsnames]{xcolor}
\newcommand{\cla}[1]{{#1}}

\begin{document}

\title{Small-scale magnetic fields of the spectroscopic binary T Tauri stars V1878 Ori and V4046 Sgr}
\titlerunning{Magnetic fields of binary T Tauri stars}

   \author{A. Hahlin \and O. Kochukhov}
   
   \institute{Department of Physics and Astronomy, Uppsala University, Box 516, SE-751 20 Uppsala, Sweden \\\email{axel.hahlin@physics.uu.se}}

    \date{Received: 13-10-2021, Accepted: 05-01-2022}
    
\abstract
{}
{The goal of this study is to investigate the small-scale magnetic fields of the two spectroscopic binary T Tauri stars V1878 Ori and V4046 Sgr. This is done to complete the observational characterisation of the surface magnetic fields of these stars because only their large-scale magnetic fields have been studied with Zeeman Doppler imaging (ZDI) so far.}
{To investigate the small-scale magnetic fields, the differential Zeeman intensification of near-infrared \ion{Ti}{i} lines was investigated using high-resolution archival spectra obtained with the ESPaDOnS spectrograph at the CFHT. In order to study the binary components separately, the spectra were disentangled by considering observations taken at different orbital phases. The Zeeman-intensification analysis was performed based on detailed polarised radiative transfer calculations aided by the Markov chain Monte Carlo inference, treating magnetic field filling factors and other stellar parameters that could affect the spectra as free parameters.}
{The obtained average magnetic field strengths of the components of V1878 Ori are $1.33\pm0.08$ and $1.57\pm0.09$~kG, respectively. Previous ZDI studies of V1878 Ori recovered about 14 and 20\% of this total magnetic field strength. For V4046 Sgr, the magnetic field strengths are $1.96\pm0.18$ and $1.83\pm0.18$~kG, respectively. In this case, about 12 and 9\% of the total magnetic field strength was detected by ZDI.}
{The small-scale magnetic field strengths obtained from Zeeman intensification are similar for the two components of each binary. This is in contrast to the large-scale magnetic fields obtained from ZDI investigations, performed using the same observations, which found that magnetic field strengths and topologies vary significantly in the components of the two binaries. While the large-scale field might look significantly different, the overall magnetic energy, primarily carried by the small-scale magnetic fields, appears to be quite similar. This indicates that the efficiency of the magnetic dynamo is comparable for the components of the two binaries.} 

\keywords{binaries: spectroscopic -- stars: activity -- stars: magnetic field -- stars: variables: T Tauri -- techniques: spectroscopic}
\maketitle
%
\section{Introduction}

Magnetic fields are important during the early stages of star formation. During the pre-main-sequence, accretion of matter from the surrounding disk onto the star occurs along magnetic field lines \citep{hartmann:2016}. This accretion is accompanied by a corresponding outflow that is required to explain the angular momentum loss that is associated with young stars. This mass outflow was predicted by \cite{matt:2005} to correspond to around 0.1 of the mass accretion for sufficient angular momentum to be lost to explain the rotation rates of young stars. Observations made by \cite{watson:2016} confirmed this theory, finding average outflow accretion ratios close to expected values. The magnetic fields play a role in driving the stellar wind through Alfve\'n waves \citep{holst:2014}. Magnetic fields are also connected to several forms of activity phenomena, such as stellar X-ray emission \citep{pevtsov:2003}. These properties will affect the surrounding environment, such as exoplanets or disks. To understand these aspects of stars, detailed properties of their magnetic fields are needed.

A common method for studying the magnetic fields of stars is Zeeman Doppler imaging \citep[ZDI, ][]{donati:2009,kochukhov:2016}. This method allows reconstructing a vector magnetic field structure on the surface of a rotating star \cla{from spectropolarimetric time-series observations}. While the information obtained from these studies provides valuable information about the global magnetic properties, \cla{ZDI does not probe the smaller spatial scales on the stellar surface because the spatial resolution is limited. This limitation is caused by magnetic flux cancellation of nearby surface elements with opposite magnetic polarity, which significantly reduces the observed disk-integrated polarisation signal. For this reason, ZDI is incapable of studying magnetic structures below this limit.} A complementary approach to measuring the magnetic field is to use the Zeeman broadening \citep[see e.g.][]{reiners:2012} by observing changes in the profiles of magnetically sensitive spectral lines. This method does not suffer from the small-scale magnetic signal cancellation that plagues ZDI and provides robust average properties of the magnetic field over the entire surface of late-type stars. Multiple studies comparing results from the two magnetic diagnostic methods have shown that most of the magnetic energy on the surfaces of cool stars is carried within these small-scale magnetic fields \citep{see:2019,lavail:2019,kochukhov:2020a}. For this reason, conclusions about the magnetic field of a star based on the large-scale structure alone are premature. 

An application of Zeeman broadening relies on the magnetic field being a dominant contributor to the spectral line broadening. The rapid rotation of active young stars means that magnetic broadening is challenging to detect because the magnetic broadening is smeared out by rotational Doppler broadening. In addition, magnetic broadening is also strongly wavelength dependent, making near-infrared observations preferable. Another property of the magnetic fields can be used in this case, specifically, that magnetic fields will increase the equivalent width of magnetically sensitive lines. This is caused by the desaturation of the lines that occurs due to the Zeeman splitting, effectively allowing more of the stellar flux to be absorbed by the line. This effect is known as Zeeman intensification \citep[e.g.][]{kochukhov:2017,kochukhov:2019}. This approach can be used to obtain information about the magnetic field even for very rapidly rotating stars and using observations at lower wavelengths than typically required for a Zeeman-broadening analysis \citep[see e.g.][]{hahlin:2021}.

In this magnetic field study, two interesting spectroscopic binaries were selected for investigation with the Zeeman-intensification technique, V1878 Ori (Parenago 523, RX J0530.7-0434, 1RXS J053043.1-0434553) and V4046 Sgr (HDE 319139). These binaries are particularly useful because their component masses are similar. The components of V1878 Ori have essentially the same projected mass ($M\sin^3{i}$) of $1.54\,M_\sun$ \citep{covino:2001,lavail:2020}. Similarly, V4046 Sgr has component masses of $0.91$ and $0.87\,M_\sun$ \citep{stempels:2004}. The close similarity between the components in the two binaries and the fact that binary components form simultaneously \citep{raghavan:2010} means that the conditions within the components should be similar. The magnetic dynamo is thought to be dependent on only a few basic stellar parameters, such as mass and age \citep{brun:2017}, therefore the magnetic field generation should be similar in the two components of these systems. However, previous ZDI results contradicted this prediction, which might indicate that dynamo processes are different even in almost identical stars. However, as most magnetic energy in late-type stars is carried in small-scale magnetic fields, ZDI might not be the most appropriate method to assess the efficiency of the magnetic dynamo. For this reason, results from a Zeeman-broadening or -intensification analysis could provide additional observational constraints for a comparison of the magnetic field generation in stars with otherwise similar properties.

V1878 Ori is a binary with an eccentric orbit. The two components are intermediate-mass T Tauri stars (IMTTS). IMTTS are progenitors of Herbig Ae/Be stars and eventually A/B-type main-sequence stars. Few of these stars are considered magnetically active;\cla{ only a few Herbig Ae/Be} have detectable magnetic fields \citep{alecian:2013}. 
\cla{In their investigation of the incidence of magnetism among IMTTS, \citet{villebrun:2019} identified a transition region in which the magnetic detection probability shifts from close to 100\,\% to about 10\,\%. This occurs when the convective zone is reduced to a mass lower than $\sim$\,2\,\% of the total mass of the star. The magnetic fields of
} V1878 Ori have previously been investigated by \citet{lavail:2020}. The authors reconstructed the global surface magnetic field maps of the two components. They also performed X-ray observations to determine whether any variation in X-ray activity between periastron and apastron could be observed. The aim was to investigate the extent of magnetospheric interaction during periastron because \cite{getman:2016} found tentative evidence of an increased X-ray activity near periastron. An increased activity could not be confirmed by \citet{lavail:2020}, indicating little interaction between the two magnetospheres. The ZDI maps obtained by \citet{lavail:2020} showed significantly different field strengths and topologies of the large-scale fields of the two components. In particular, the average magnetic field strength of the secondary was found to be almost twice as strong as that of the primary component.

V4046 Sgr contains two T Tauri stars with masses very similar to that of the Sun. Therefore, these stars are progenitors to a pair of Sun-like stars. The magnetic fields of the components of V4046 Sgr have also previously been investigated with ZDI by \citet{donati:2011}. Similarly to the results for V1878 Ori, the magnetic field maps of V4046 Sgr show a large difference in the field strengths and topologies for the two components. The ZDI maps of both stars have been extrapolated out into the magnetosphere by \citet{gregory:2014}, showing connected field lines between the two components. \cite{stempels:2004} found periodic variability of emission lines. This was interpreted as a co-rotating gas. Similar flux variability was discovered by \cite{argiroffi:2012} as they studied the system with X-ray observations. This finding supported the magnetically driven accretion scenario as the gas from the circumstellar disk falls down onto the stars along the magnetic field lines.

Because these two systems have already been investigated with ZDI, V1878 Ori by \cite{lavail:2020} and V4046 Sgr by \cite{donati:2011}, the large-scale magnetic properties are already well known. In order to complete the picture of the magnetic field, an investigation of the small-scale magnetic field properties is needed. 

In Sect.~\ref{sec:Obs} the spectroscopic observations we used in our study of the two binary system are discussed. Section~\ref{sec:SB} presents the procedure of obtaining disentangled, time-averaged spectra of the individual components. Section~\ref{sec:ZI} discusses the method of extracting information about the magnetic field and determines other relevant stellar parameters. The results of the Zeeman-intensification study of the two components in each binary are also presented in this section. Finally, the overall results of our study are discussed in Sect.~\ref{sec:summary}.

\section{Observations}
\label{sec:Obs}
Observations of both stars were obtained at the Canada-France-Hawaii Telescope (CFHT) using the high-resolution spectrograph ESPaDOnS \citep{donati:2003,wade:2016}. This instrument has a resolving power of 65 000 and covers the wavelength range from 3700 to 10000\,\AA. 

The observations of V1878 Ori were obtained during two periods in 2016, the first between January 14 and 29, and the second between March 25 and 29. These data were collected in the context of the Binarity and Magnetic Interactions in various classes of Stars \citep[BinaMIcS,][]{alecian:2015} large programme. In total, 25 observations of V1878 Ori are available. A total of 7 observations of V4046 Sgr were obtained in 2009 between September 3 and 9. 

The observations of the two binaries are the same as were employed in the ZDI analyses of the global magnetic field topologies of these systems by \citet{lavail:2020} for V1878 Ori and \citet{donati:2011} for V4046 Sgr. That we use the same observations will allow a direct comparison between the small- and large-scale magnetic fields at the same point in time for both systems.

The ESPaDOnS spectra of the two targets were reduced using the UPENA pipeline running the LIBRE-SPIRIT software \citep{donati:1997}. The pipeline data products were downloaded from the CFHT Science Archive hosted by the Canadian Astronomy Data Centre\footnote{\url{https://www.cadc-ccda.hia-iha.nrc-cnrc.gc.ca/en/cfht/}}. The median signal-to-noise ratio (S/N) of the reduced spectra was estimated for the spectral order centred at 9840~\AA, closest to the spectral lines of interest for the Zeeman-intensification analysis. The observations of V1878 Ori have a median S/N of 175, while V4046 Sgr data have a median S/N of 72. A complete list of observations and their S/N can be found in Appendix~\ref{appena} (Tables~\ref{tab:OriSN} and \ref{tab:SgrSN}).

\section{Spectral disentangling}
\label{sec:SB}

\begin{table}
\caption{Orbital solutions used for V1878 Ori and V4046 Sgr.}
\label{tab:OrbSol}
\begin{tabular}{lcc}
\hline \hline
 & V1878 Ori$^{\mathrm{a}}$ & V4046 Sgr$^{\mathrm{b}}$ \\
\hline
$P_{\mathrm{orb}}$ (d) & $40.58318\pm0.00025$ & 2.4213459 \\
$T_{0}$ (HJD) & $2451098.925\pm0.065$ & 2446998.335 \\
$\gamma$ (km\,s$^{-1}$) & $32.919\pm0.066$ & $-6.94$ \\
$K_{\mathrm{A}}$ (km\,s$^{-1}$) & $47.08\pm0.18$ & 54.16 \\
$K_{\mathrm{B}}$ (km\,s$^{-1}$) & $47.74\pm0.13$ & 56.61 \\
$e$ & $0.3157\pm0.0018$ & 0.00 \\
$\omega$ (deg) & $287.45\pm0.51$ & 0.00\\
\hline
\end{tabular}
\tablefoot{(a) \cite{lavail:2020}, (b) \cite{stempels:2004}}
\end{table}

Because double-line
spectroscopic binaries show a complicated and time-dependent line blending, the spectra of their components need to be disentangled, or separated, to study the components individually. An additional element that needs to be accounted for when interpreting observations in the near-infrared wavelength region is the strong telluric absorption. Here we applied the spectral separation procedure similar to the disentangling performed for other binary systems that have previously been investigated using Zeeman intensification \citep{kochukhov:2019,hahlin:2021}. Because an orbital solution has already been calculated for the two binary systems \citep[][see Table~\ref{tab:OrbSol}]{lavail:2020,stempels:2004}, the predicted radial velocity values were used to initiate the disentangling process. The spectral separation procedure was carried out by fitting all available observations simultaneously with a superposition of three spectral components. The first two spectral contributions correspond to the binary components of the system, and the third component represents the telluric absorption. 
The stellar spectra were shifted according to individual radial velocities and were assumed to be constant. The telluric spectrum was shifted according to the known radial velocity difference of the observatory relative to the heliocentric reference frame and was allowed to vary in intensity to better match individual observations.
This composite spectral model was iterated until convergence was achieved, yielding disentangled spectra for both the stellar components and the telluric absorption.

The wavelength region between 9635 and 9816~\AA\ was used for the disentangling process. This wavelength range contains a set of \ion{Ti}{i} lines that is useful for Zeeman intensification. These lines are described in more detail in Sect. \ref{sec:ZI}.

\begin{figure}[!t]
\centering
\includegraphics[width=\hsize]{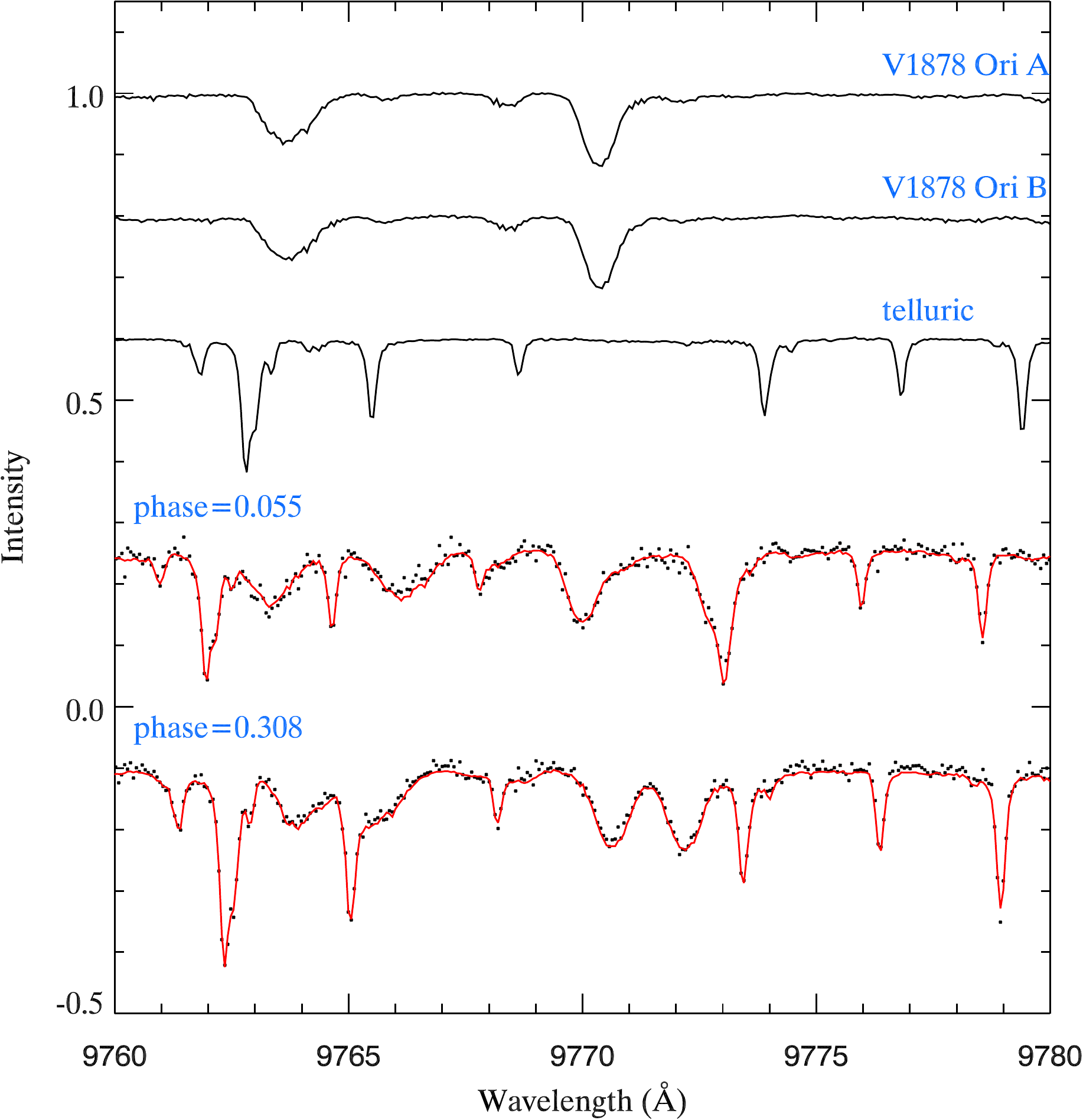}
\caption{Illustration of spectral disentangling results for V1878 Ori. The mean spectra of each component are shown at the top, followed by the derived telluric absorption (the third spectrum). The model composite spectra (solid lines) and observations (symbols) at the orbital phases close of the largest and smallest radial velocity separation of the components are compared below. The spectra are shifted vertically for visibility.}
\label{fig:DisVOri}
\end{figure}

\begin{figure}[!t]
\includegraphics[width=\hsize]{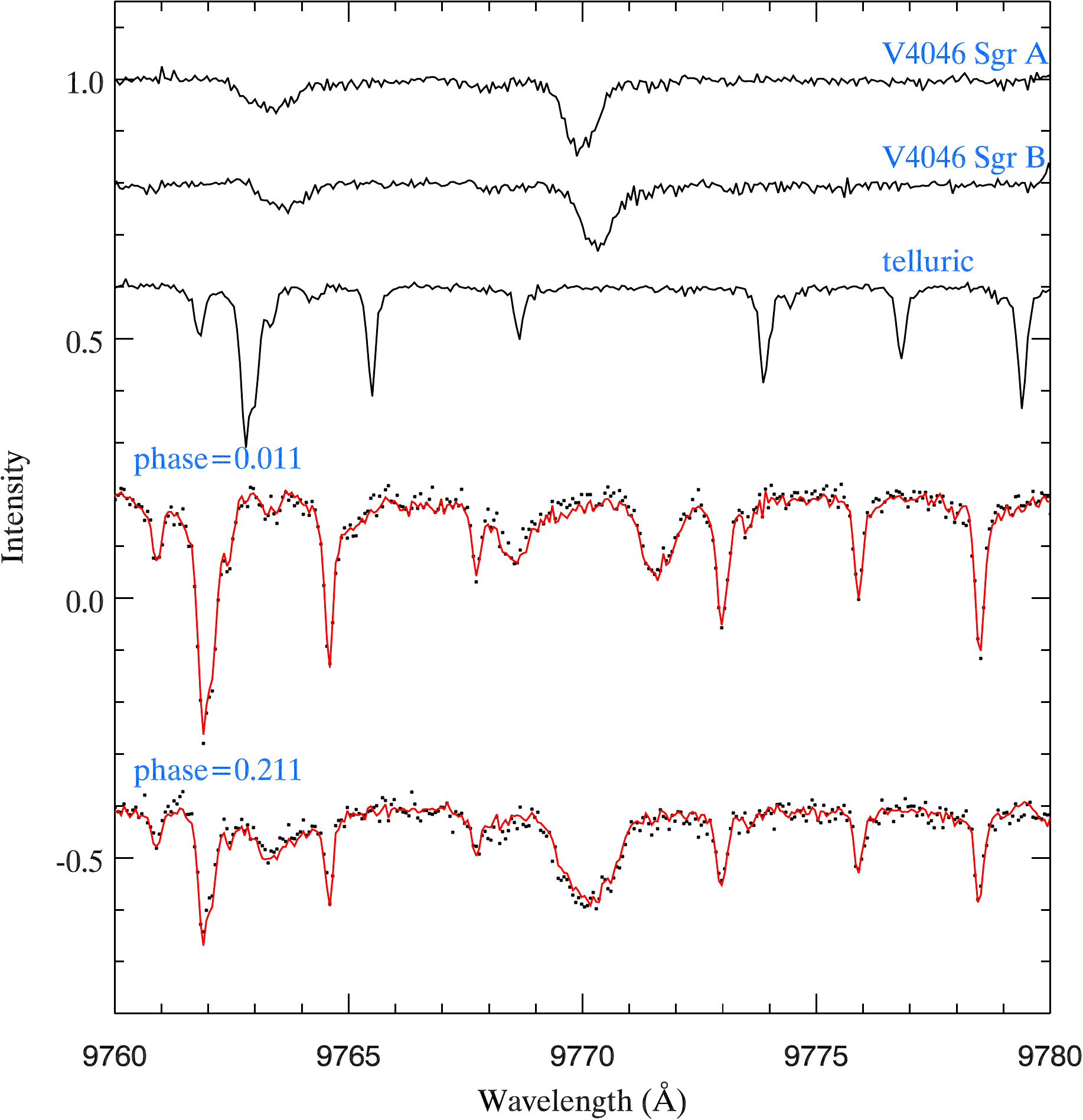}
\caption{Same as Fig.~\ref{fig:DisVOri}, but for the spectral disentangling applied to V4046 Sgr.}
\label{fig:DisVSgr}
\end{figure}

The result of this disentangling procedure is a spectrum for each star that corresponds to a mean spectrum averaged over all observations. Examples of these mean spectra are shown in Fig.~\ref{fig:DisVOri} and \ref{fig:DisVSgr} for V1878~Ori and V4046~Sgr, respectively. While it would be interesting to obtain phase-resolved intensity spectra to potentially observe variations in magnetic field strength over the rotational period, the time-dependent variability and radial velocity variation of the two components in a spectroscopic binary prevents us from obtaining spectra at each individual phase. 

Although our disentangling method is capable of extracting relative radial velocities for the two stellar components at each orbital phase, the absolute radial velocity scale remains unconstrained because no information about laboratory line positions is taken into account. If the orbital ephemeris data are inaccurate, the disentangled spectra might be shifted. This appears to be the case for the components of V4046~Sgr in Fig.~\ref{fig:DisVSgr}, where the spectral lines that should be at the same wavelength are not aligned. \citet{donati:2011} have encountered similar issues with the orbital ephemeris reported in \citet{stempels:2004} in their ZDI study of V4046~Sgr. To mitigate this issue here, we included the radial velocity offset of disentangled spectra as an additional free parameter in the spectrum synthesis modelling.

While the disentangling procedure separates the spectral lines of each component, it is not able to separate the continuum contribution of each star. This causes the absorption lines in the disentangled spectra to be shallower compared to those of a single star. In order to account for this, the luminosity ratio ($\mathrm{LR} = L_{\mathrm{A}}/L_{\mathrm{B}}$) of the system must be used to scale either the disentangled or model spectra employed to find the stellar parameters. We opted for scaling the model spectra and derived the luminosity ratio in the investigation of the magnetic field simultaneously with other parameters.

\section{Zeeman intensification}
\label{sec:ZI}
\begin{table}
  \caption{Stellar parameters employed for spectrum synthesis calculations.} 
  \label{tab:StellarPar}
  \begin{tabular}{lcccc}
    \hline \hline
     & \multicolumn{2}{c}{V1878 Ori$^{\mathrm{a}}$} & \multicolumn{2}{c}{V4046 Sgr$^{\mathrm{b}}$}\\
     & A & B & A & B \\ 
    \hline 
    $T_{\mathrm{eff}}$ (K) & 4800 & 4750 & 4370 & 4100 \\
    $\log g$ (cm\,s$^{-2}$) & 4.07 & 3.84 & 4.0 & 4.0 \\
    $v_{\mathrm{mic}}$ (km\,s$^{-1}$) & 1.0 & 1.0 & 1.0 & 1.0 \\
    \hline
  \end{tabular}
  \tablefoot{(a) \cite{lavail:2020}. (b) \cite{stempels:2004}}
\end{table}

The key advantage of Zeeman intensification over Zeeman broadening is that the former magnetic field measurement method concerns itself with only the change in equivalent width of the observed lines and does not attempt to draw information from line profile shapes. This removes the strong preference of observing at longer wavelengths inherent to Zeeman broadening. As a consequence, it becomes easier to use optical spectrographs, which are commonly used for ZDI studies, to perform a simultaneous analysis of the mean magnetic field with the data collected for global field mapping. It also removes the requirement that rotational broadening should be low so that it does not hide the magnetic broadening, allowing investigations to be performed even for rapidly rotating active stars using observations with a lower resolution and S/N. A disadvantage with primarily focusing on the equivalent width is that the magnetic effect becomes hard to distinguish from other parameters that modify the line depth, such as the element abundance. This means that magnetically insensitive lines, preferably of the same ion, must be modelled in order to determine non-magnetic parameters simultaneously with the derivation of the magnetic field.

Our study of Zeeman intensification follows previous investigations of the mean magnetic field modulus in G, K, and M stars \citep{lavail:2019,kochukhov:2020a,hahlin:2021}. The SoBAT library for IDL \citep{anfinogentov:2021} was used to perform a Markov chain Monte Carlo (MCMC) sampling in order to find a set of parameters that best reproduces observations and establish realistic confidence intervals accounting for possible correlations between parameters. In order to perform this procedure, synthetic spectra were calculated. This was done by collecting line lists from VALD \citep{ryabchikova:2015}, assuming solar metallicity from  \cite{asplund:2009}, and using MARCS \citep{gustafsson:2008} model atmospheres in combination with the polarised radiative transfer code Synmast \citep{kochukhov:2010}. Similar to most previous mean magnetic field studies of cool stars, a uniform radial magnetic field was assumed. \cla{This radial assumption will produce a wide range of different magnetic field orientations with respect to the line of sight \citep{yang:2011}. As a consequence, any disk-integrated spectrum will contain contributions from several different field directions, giving a balanced contribution to the synthetic spectra. \citet{shulyak:2014} and \citet{kochukhov:2021} showed that replacing a radial field with a horizontal field has little effect on the shape and strength of lines in intensity spectra. For this reason, it is unlikely that our choice of the radial magnetic field geometry has any significant impact on the results presented below.} 

A grid of synthetic spectra spanning suitable values of Ti abundance and magnetic field strengths was produced for each binary component. The other stellar parameters, $T_{\rm eff}$, $\log g$, and the microturbulence $v_\mathrm{mic}$, were the same as those adopted in previous studies of the two binaries \citep{stempels:2004,lavail:2020}. These parameters are listed in Table~\ref{tab:StellarPar}. The exception are the values of $v_\mathrm{mic}$ for V1878 Ori. \cite{lavail:2020} obtained significantly different values of $v_\mathrm{mic}$ for the two components. Because the stars are otherwise similar, this difference is odd. Furthermore, the $v_\mathrm{mic}=2.34\pm0.58$~km\,s$^{-1}$ obtained for the primary star is anomalously high. Typical values of $v_\mathrm{mic}$ lie in the range of 1.0--1.5~km\,s$^{-1}$ for cool stars of similar $T_{\rm eff}$ \citep[see e.g.][]{jofre:2015}. A single fixed value of $v_\mathrm{mic}$ around 1.0~km\,s$^{-1}$ is frequently assumed in abundance analyses of large samples of late-type stars \citep[e.g.][]{valenti:2005,brewer:2016}. For this reason, we set $v_\mathrm{mic}$ equal to 1.0~km\,s$^{-1}$ for the two components of V1878 Ori.

Because the two binaries considered here are T Tauri stars, veiling caused by the accreting matter can significantly affect the stellar spectra by reducing the depth of spectral lines. However, in this particular case, little evidence of strong veiling has been found for either V1878 Ori \citep{lavail:2020} or V4046 Sgr \citep{stempels:2004}. For this reason, we did not include any correction for veiling in our analysis of either binary. In any case, a small veiling correction would be indistinguishable from a change in element abundance.

Ten spectral lines belonging to the same \ion{Ti}{i} multiplet formed between the a$^5$F and z$^5$F$^{\rm o}$ terms of neutral titanium were used in our Zeeman-intensification study. They have been used successfully in past studies of M dwarfs \citep[see][]{kochukhov:2017,shulyak:2017,kochukhov:2019}. These lines are located in the wavelength interval 9640--9800 \AA\ and have a range of effective Land\'e factors between 0 and 1.55. The \ion{Ti}{i} 9743.61 \AA~is a magnetic null line, meaning that it is insensitive to a magnetic field. Combining the modelling of this line with the interpretation of magnetically sensitive lines allows disentangling the intensification due to the magnetic field from non-magnetic effects. 

\begin{table}[]
    \centering
    \caption{\ion{Ti}{i} lines used for the Zeeman-intensification analysis.}
    \label{tab:Ti_lines}
    \begin{tabular}{cccccc}
        \hline \hline
        $\lambda$ (\AA) & $E_{\rm lo}$ (eV) & $E_{\rm up}$ (eV) & $\log{gf}$ & $g_{\mathrm{eff}}$ & $\Delta W/W_0^\mathrm{a}$\\
        \hline
        9647.37 & 0.8181 & 2.1030 & $-1.434$ & 1.53 & 2.64\,\%\\
        9675.54 & 0.8360 & 2.1171 & $-0.804$ & 1.35 & 0.85\,\%\\
        9688.87 & 0.8129 & 2.0922 & $-1.610$ & 1.50 & 4.98\,\%\\
        9705.67 & 0.8259 & 2.1030 & $-1.009$ & 1.26 & 0.99\,\%\\
        9728.41 & 0.8181 & 2.0922 & $-1.206$ & 1.00 & 0.84\,\%\\
        9743.61 & 0.8129 & 2.0851 & $-1.306$ & 0.00 & 0.00\,\%\\
        9770.30 & 0.8484 & 2.1171 & $-1.581$ & 1.55 & 2.85\,\%\\
        9783.31 & 0.8360 & 2.1030 & $-1.428$ & 1.26 & 2.22$^\mathrm{b}$\,\%\\
        9783.59 & 0.8181 & 2.0851 & $-1.617$ & 1.49 & 4.97$^\mathrm{b}$\,\%\\
        9787.69 & 0.8259 & 2.0922 & $-1.444$ & 1.50 & 2.53\,\%\\
        \hline
    \end{tabular}
    \tablefoot{(a) The relative change in equivalent width caused by a magnetic field strength of 1\,kG assuming $T_{\rm eff}=4750$, $\log g=4.0$, $v\sin{i}=5$~km\,s$^{-1}$, and $v_\mathrm{mic}=1$~km\,s$^{-1}$. (b) Lines are blended; the effective change in equivalent width with both lines combined is 3.34\,\%.}
\end{table}

Detailed information about the \ion{Ti}{i} lines used in our analysis is provided in Table~\ref{tab:Ti_lines}. The last column in this table quantifies the relative magnetic sensitivity by comparing the change in equivalent width caused by the introduction of a 1\,kG magnetic field. This assessment shows that the \ion{Ti}{i} 9688.87 and 9783.59\,\AA\ lines in particular have the strongest response to a magnetic field, whereas the \ion{Ti}{i} 9743.61\,\AA\ line exhibits no response at all, and a handful of lines (9675.54, 9705.67, 9728.41~\AA) show a weak response. The 9783.59\,\AA\ line, however, is blended with \ion{Ti}{i} 9783.31~\AA,\ which reduces the magnetic effect. It is also worthwhile mentioning that while the effective Land\'e factor is a good indicator of the magnetic sensitivity of spectral lines, it is not always sufficient because Zeeman splitting pattern also plays a role. For example, the \ion{Ti}{i} 9770.30\,\AA\ line has the largest effective Land\'e factor of the transitions, but not the strongest response to the presence of a magnetic field.

In a change from our previous investigations of Zeeman intensification in the spectra of late-type active binaries \citep{kochukhov:2019,hahlin:2021}, the inference for the two binary components was performed simultaneously. This is justified by the fact that binaries form together from the same molecular cloud. For this reason, the surface chemical abundances of binary components with a similar mass can be assumed to be the same. By performing the inference simultaneously, these abundances can be constrained by observations of both stars instead of one at a time. In the particular case of V1878 Ori, the metallicities of its components were estimated by \cite{lavail:2020}. While a slight difference between the two components was obtained, it was not significant given the uncertainties. This indicates that the Ti abundance of the two stars could be approximated by a single value.
The magnetic field is parametrised with a set of \cla{$N$} filling factors $f_i$, representing fractions of the stellar surface covered with a magnetic field of a certain strength $B_i$,
\begin{equation}
S_{\rm tot} = \sum_{i=1}^N f_i S(B_i).    
\end{equation}
Our model is capable of handling any number of magnetic filling factors. For this work, the filling factors are defined in steps of 2\,kG, starting at 0\,kG. In order to ensure that the magnetic filling factors are properly normalised, they are also constrained according to the equation
\begin{equation}
\label{eq:ff-constraint}
\sum_{i=1}^N f_i=1.
\end{equation}
This prevents the filling factors from covering more or less than the entire star; this solution would not be physical. Because the number of magnetic field filling factors is not known a priori, it is determined by comparing the Bayesian information criterion \citep[BIC,][]{sharma:2017}, 
\begin{equation}
  \label{eq:BIC}
  \mathrm{BIC}=-2\ln(Y|\hat{\theta})+d\ln{n},
\end{equation}
obtained for inferences with different number of filling factors. Here $\ln{(Y|\hat{\theta})}$ represents the likelihood of the best fit for the spectra $Y$ using the parameters $\hat{\theta}$, $d$ is the number of data points in the observations for which the inference is performed, and $n$ is the number of parameters used in the inference. This equation weights the accuracy of the fit against the number of parameters used in the inference. The model with the lowest BIC value is selected as the most suitable model for each binary system. \cla{This model is found by adding an increasing number of magnetic field components to the model until the BIC value begins to increase, indicating that no further improvement can be made by increasing the complexity of the model.}

Because Zeeman intensification is primarily focused on the increase in equivalent width, the magnetic field diagnostic carried out here becomes entangled with other parameters. This means that the inference also includes a determination of the Ti abundance (common for the two binary components) and individual projected rotational velocities $v\sin{i}$. Because the stars are spectroscopic binaries, the (common) light ratio parameter is also included in order to scale the model spectra to the disentangled spectra according to the following formulas: 
\begin{equation}
    \label{eq:LR_scale}
    S_{\mathrm{scaled}}^{\mathrm{A}}=
    \frac{S_{\mathrm{syn}}^{\mathrm{A}}}{1+1/\mathrm{LR}}+\frac{1}{1+\mathrm{LR}}
\end{equation}
for the primary, and 
\begin{equation}
    \label{eq:LR_scale2}
    S_{\mathrm{scaled}}^{\mathrm{B}}=\frac{S_{\mathrm{syn}}^{\mathrm{B}}}{1+\mathrm{LR}}+\frac{\mathrm{LR}}{1+\mathrm{LR}}
\end{equation}
for the secondary. The individual radial velocities are also included to compensate velocity offsets that appear in the disentangling.

In order to ensure that the MCMC walker had sufficient time to find the maximum likelihood region of the parameter space, a burn-in length of 20\,000 steps was used. The sampling continued until it reached convergence. This was determined by calculating the effective sample size \citep[ESS;][]{sharma:2017} from the autocorrelation time of the walker. The requirement for convergence was set to be when the the ESS reached 2000. This convergence was tested every 10\,000th step of the walker. The values reported for each star are median values calculated over the obtained posterior distributions. Reported uncertainties correspond to a confidence level of 68\%.  

\subsection{V1878 Ori}

\begin{figure*}
\centering
\includegraphics[width=0.49\hsize]{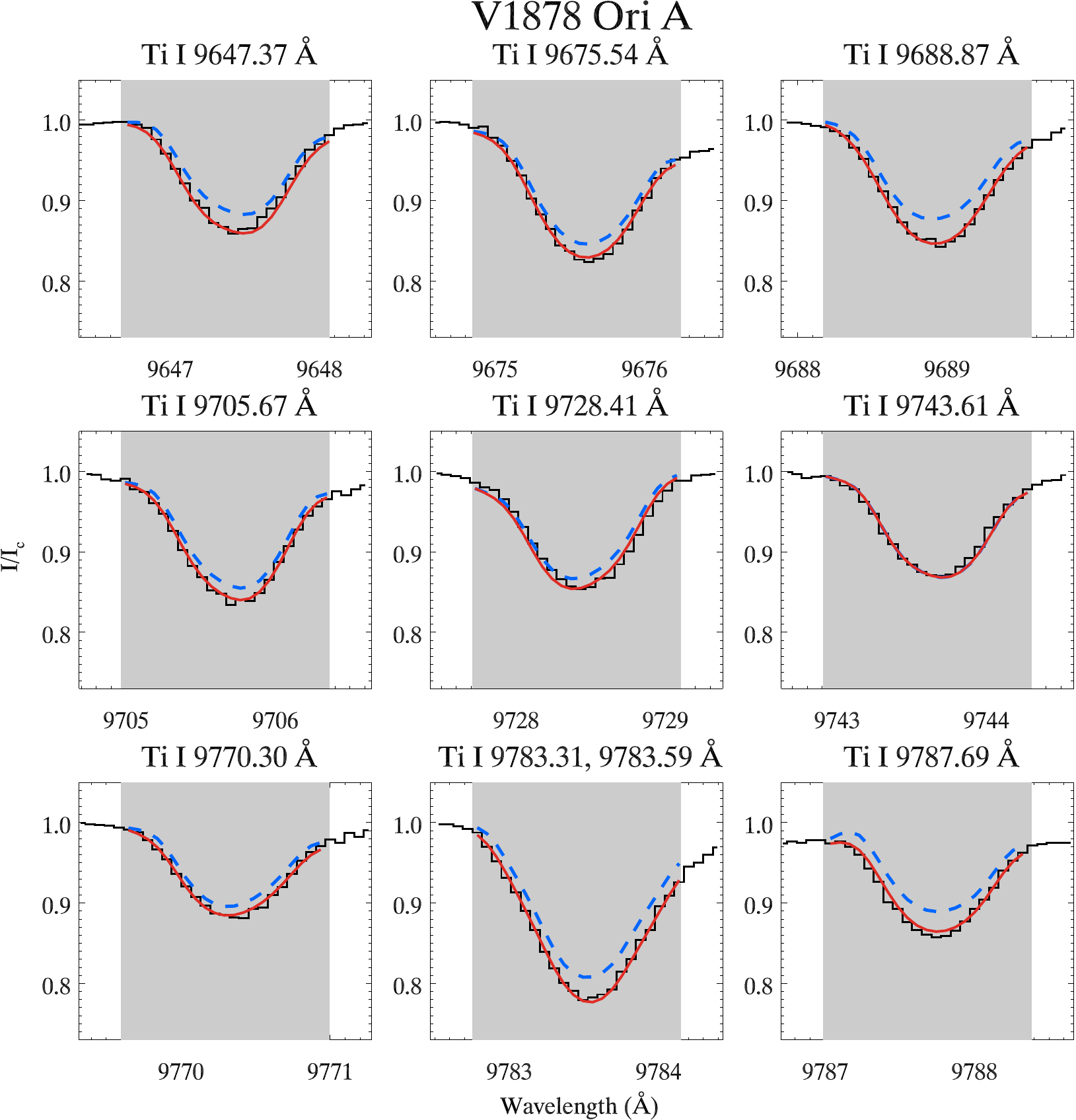}
\includegraphics[width=0.49\hsize]{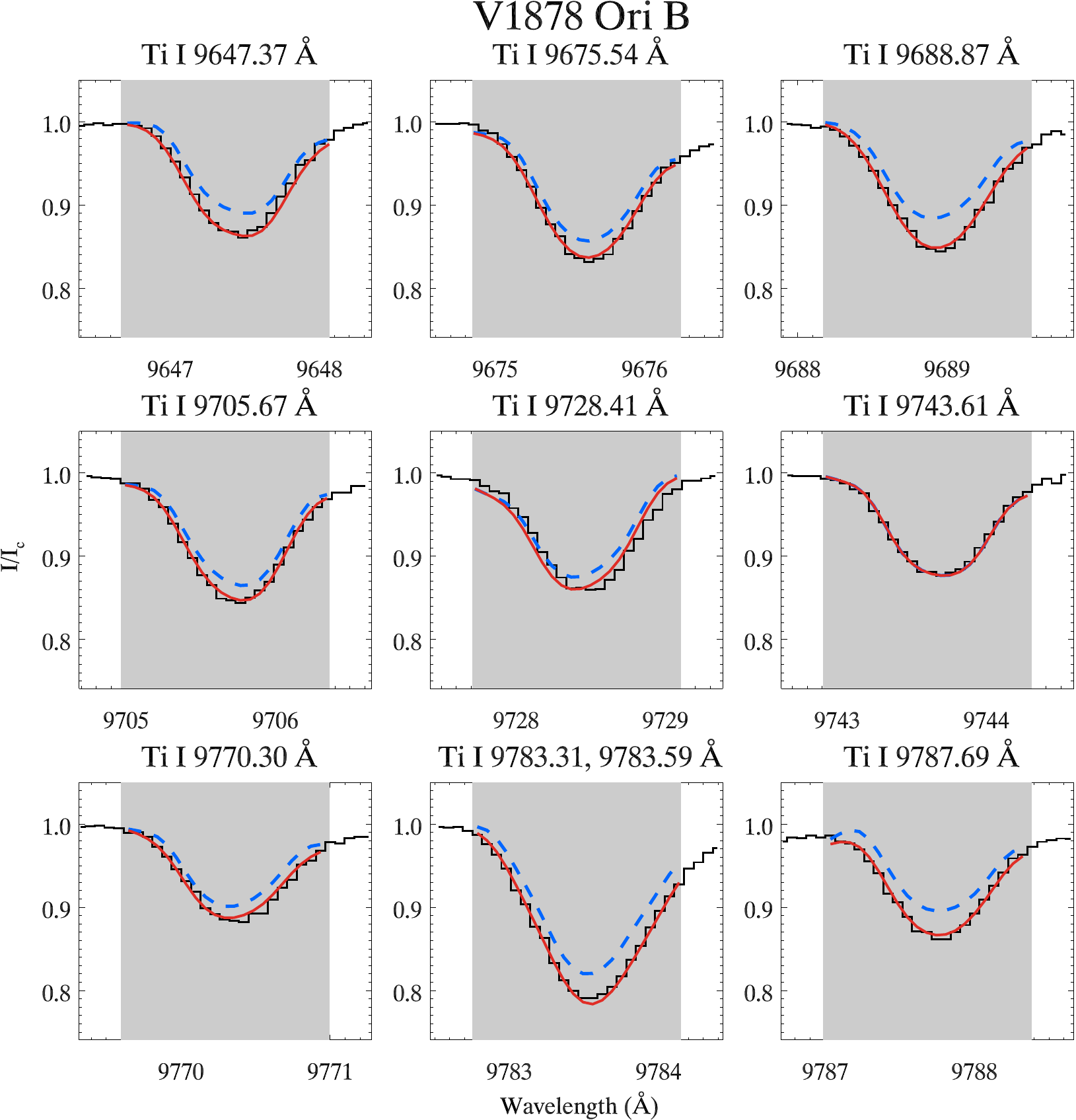}
\includegraphics[width=0.49\hsize]{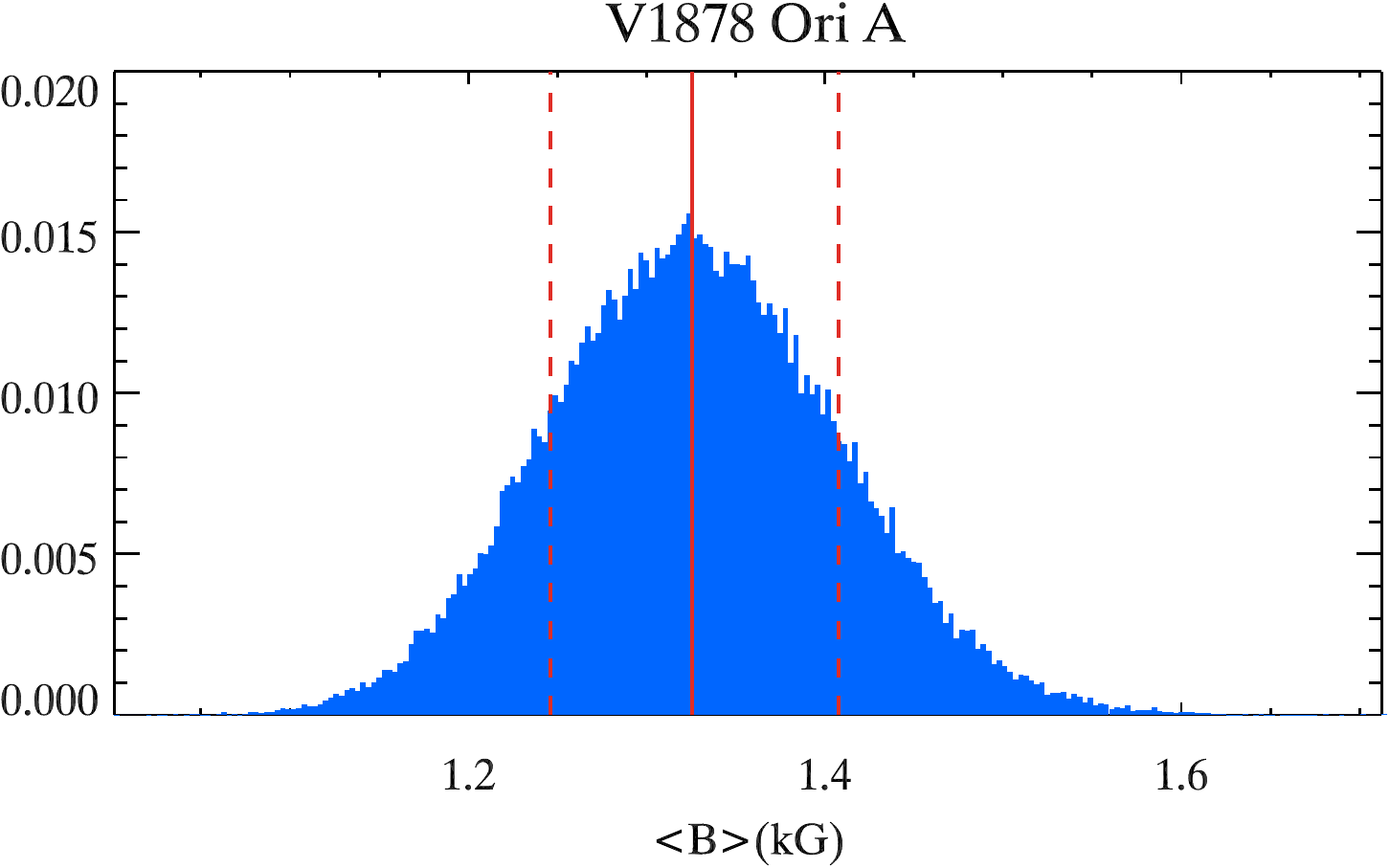}
\includegraphics[width=0.49\hsize]{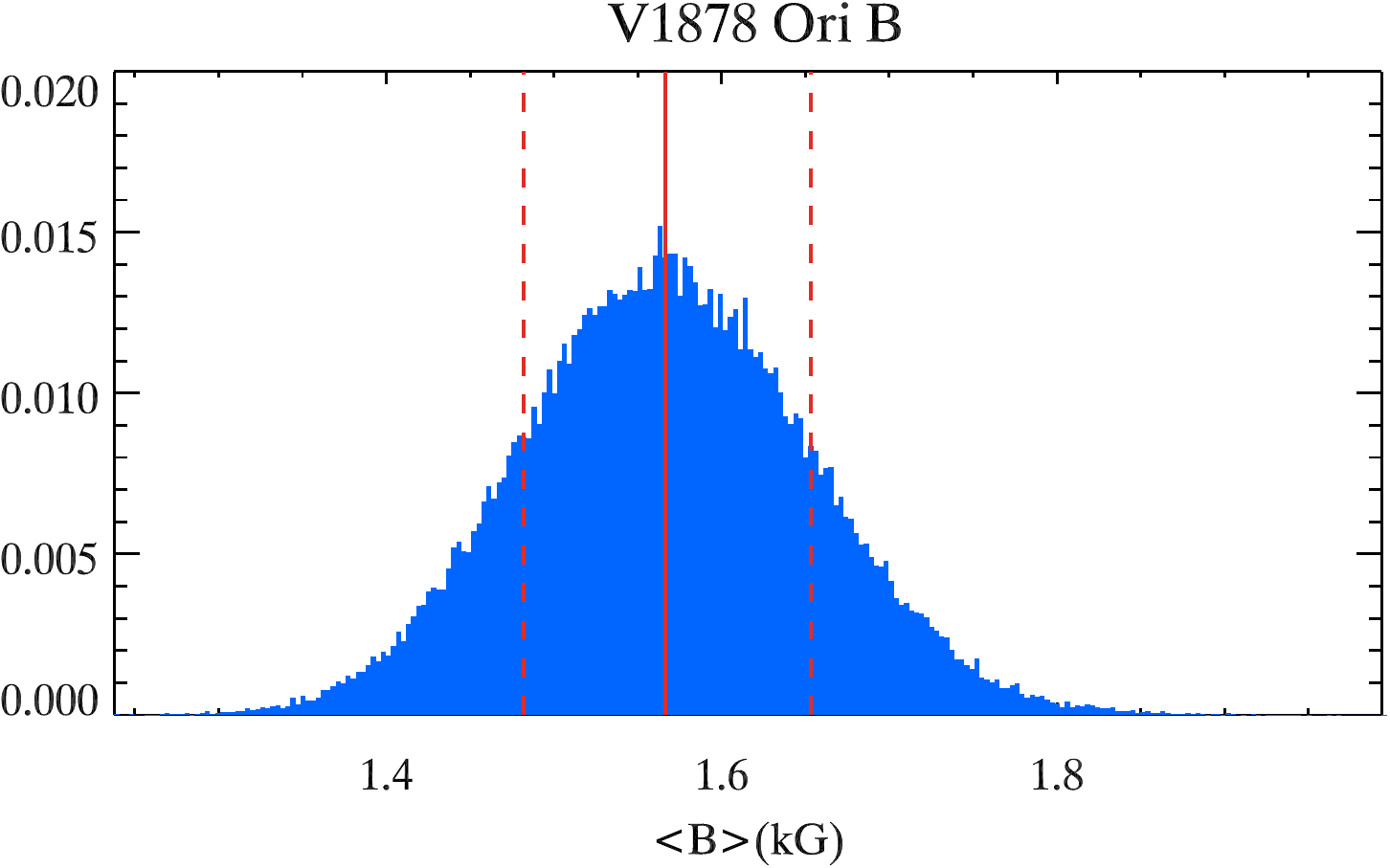}
\caption{Inference result for V1878 Ori \textit{Top.} Fit to the observed spectral lines (black histograms) with the spectra corresponding to the best-fitting parameters (solid red line). The dashed blue lines correspond to theoretical spectra without accounting for the magnetic field. 
Central wavelengths in \AA\ of each line are given above the corresponding panels. 
\textit{Bottom.} Posterior distribution of the average surface magnetic field strength in the components of V1878 Ori. The vertical lines show the median (solid line) value and 68\% percentile (dashed lines).}
\label{fig:OriMagnetic}
\end{figure*}

The priors for the parameter inference were selected to provide a generous range so as to not interfere with the magnetic field determination. The Ti abundance was given a uniform prior between $-6.90$ and $-7.40$ to cover the uncertainties from the two values determined by \cite{lavail:2020}. The luminosity ratio in previous studies was assumed to be equal to 1. Here we set a uniform prior of $1\pm0.5$. With the $v\sin{i}$ of the two components having similar values, they were both given identical uniform priors corresponding to $15\pm2.5$~km\,s$^{-1}$. All magnetic field filling factors were given uniform priors between 0 and 1, with the additional constraint defined by Eq.~(\ref{eq:ff-constraint}). As discussed in Sect.~\ref{sec:SB}, the lines in disentangled spectra might be slightly shifted from their expected laboratory positions. Even if there is no obvious shift for either component of V1878 Ori, the radial velocities of the two components were allowed to vary in a range of $\pm2$~km\,s$^{-1}$ in order to ensure that the shift did not affect the final result.

The magnetic field was studied using models with a different number of magnetic filling factors\cla{. The model chosen was the model with lowest BIC value as calculated from Eq.~(\ref{eq:BIC}). The obtained BIC values for all models are listed in Table \ref{tab:BIC}}. For V1878 Ori, the optimal number of parameters was a total of three filling factors for each component, corresponding to field strengths of 0, 2, and 4~kG, respectively. The resulting parameters from the inference are listed in Table~\ref{tab:OriResult}. The best fit to spectral lines and the posterior distribution of the average magnetic field is presented in Fig. \ref{fig:OriMagnetic}. The complete inference result in the form of posterior distributions of the selected parameters is shown in the appendix, Fig. \ref{fig:OriResults}. \cla{Some inference parameters are correlated with each other. The filling factors of different magnetic field strengths are connected because the intensification is primarily sensitive to the average magnetic field strength. The filling factors and $v\sin{i}$ are also correlated. Because the correlation between $v\sin{i}$ and $f_{4}$ is negative and that between $v\sin{i}$ and $f_{2}$ is negative, it indicates an interplay between magnetic and rotational broadening in these \ion{Ti}{i} lines.} The mean magnetic field strength of V1878 Ori B is slightly higher than that of V1878 Ori A: $1.57\pm0.09$ versus $1.33\pm0.08$~kG.

The two overall metallicities [M/H] determined by \citet{lavail:2020} were $0.06\pm0.17$ and $-0.07\pm0.26$ for V1878 Ori A and B, respectively. Assuming Ti follows this general trend, these values correspond to Ti abundances of $-7.03$ and $-7.16$ in the $\log(N_{\rm Ti}/N_{\rm tot})$ units, respectively. Our joint Ti abundance agrees with these two values within the uncertainties. This indicates that the titanium abundances in the two components in the V1878 Ori system can be described by a single value. The other parameters obtained through the inference also agree with the values previously determined by \citet{lavail:2020}.

\begin{table}
\caption{Obtained parameters for the components of V1878 Ori.}
\label{tab:OriResult}
\begin{tabular}{lrr}
\hline \hline
  Parameter & \multicolumn{1}{c}{A} & \multicolumn{1}{c}{B} \\
\hline
    $\langle B_I \rangle$ (kG) & $1.33\pm0.08$ & $1.57\pm0.09$\\
    $f_0$ & $0.50\pm0.04$ & $0.37\pm0.04$ \\
    $f_2$ & $0.34\pm0.07$ & $0.47\pm0.07$ \\
    $f_4$ & $0.16\pm0.04$ & $0.16\pm0.04$ \\
    $v\sin{i}$ (km\,s$^{-1}$) & $14.14\pm0.14$& $13.32\pm0.15$\\
    $\varepsilon_{\mathrm{Ti}}$ & \multicolumn{2}{c}{$-6.93\pm0.02$} \\
    LR & \multicolumn{2}{c}{$1.08\pm0.02$} \\
    \hline
\end{tabular}
\tablefoot{$f_0$ is determined from eq. \ref{eq:ff-constraint}}
\end{table}
\subsection{V4046 Sgr}

\begin{figure*}
\centering
\includegraphics[width=0.49\hsize]{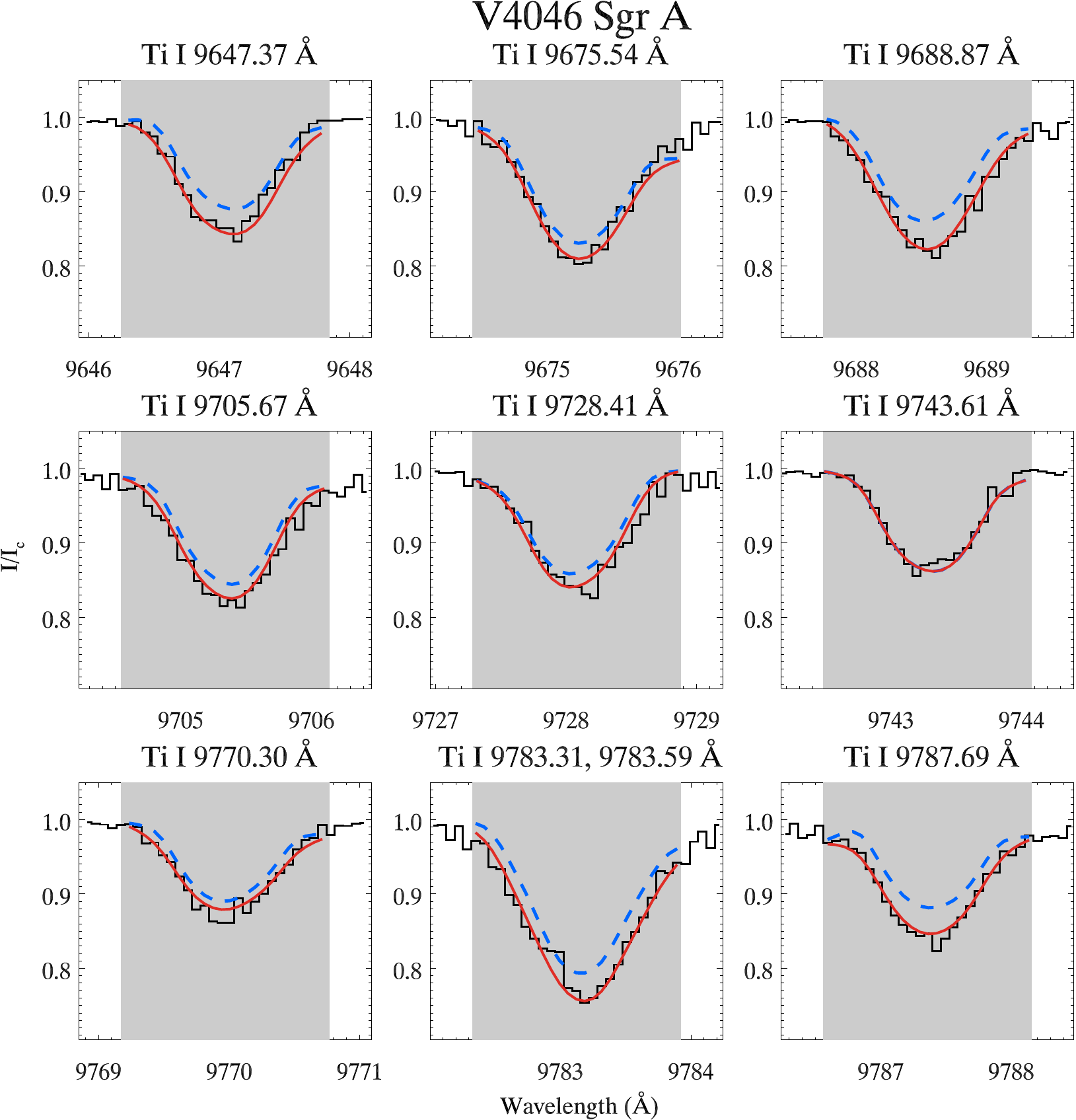}
\includegraphics[width=0.49\hsize]{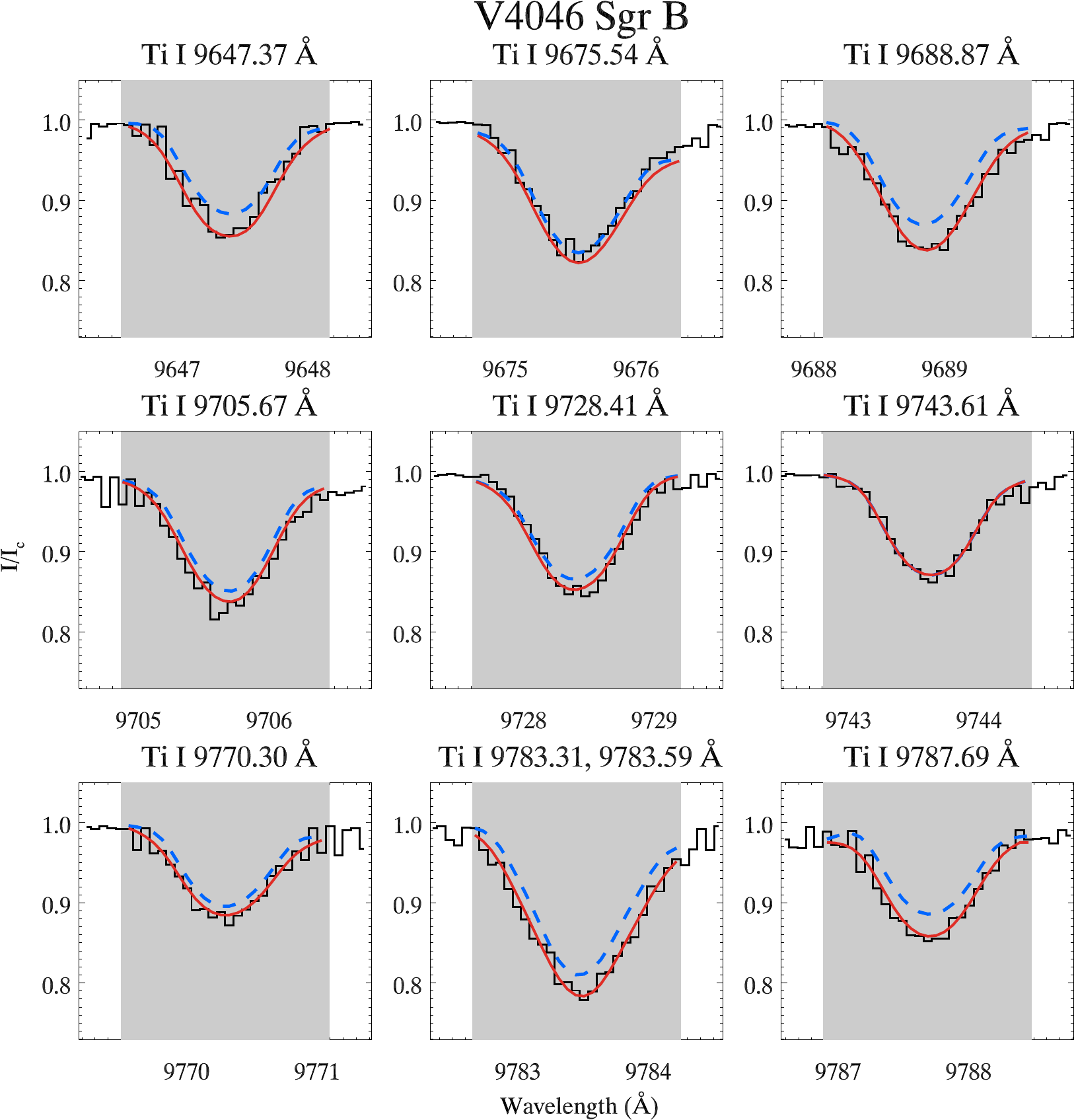}
\includegraphics[width=0.49\hsize]{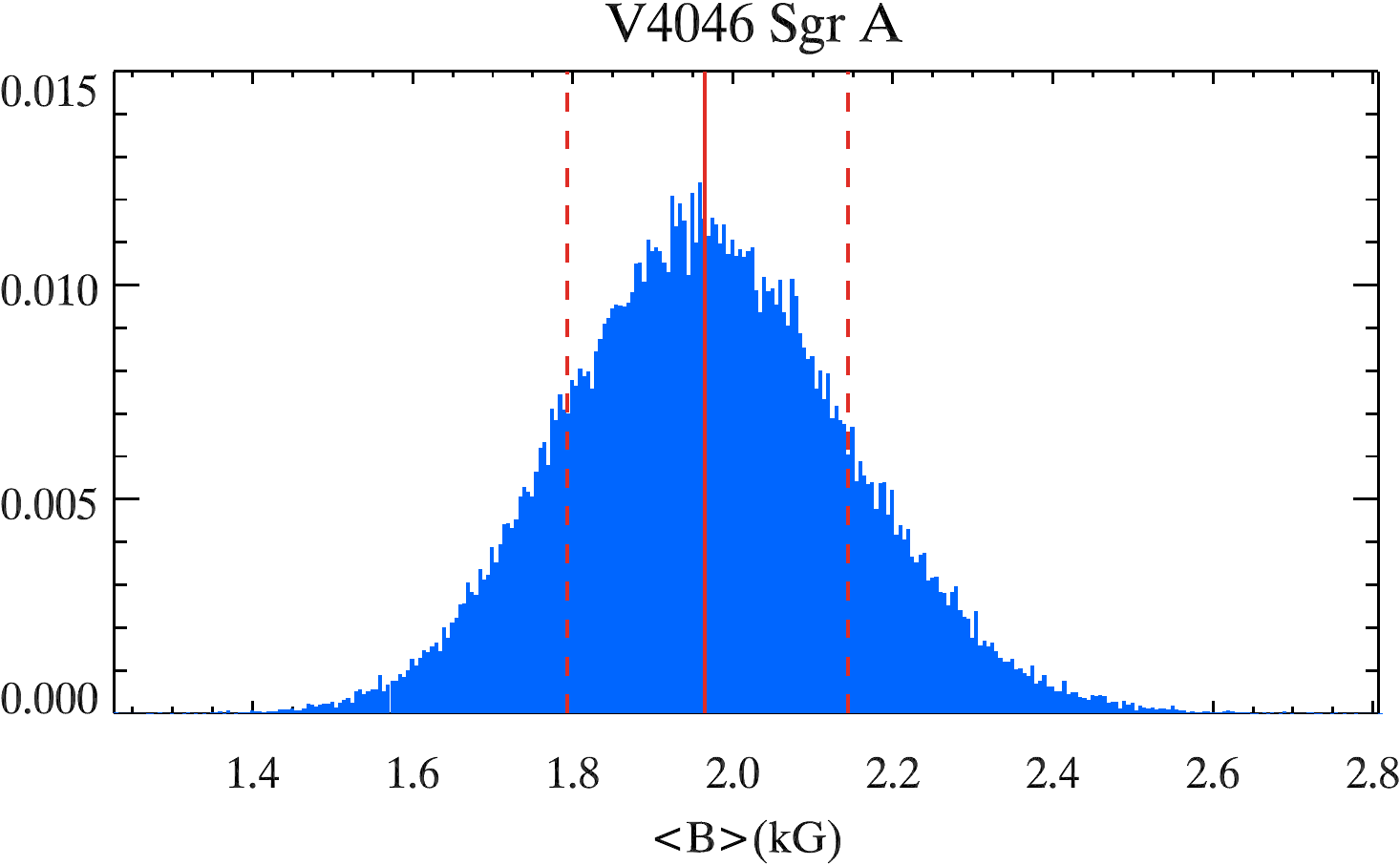}
\includegraphics[width=0.49\hsize]{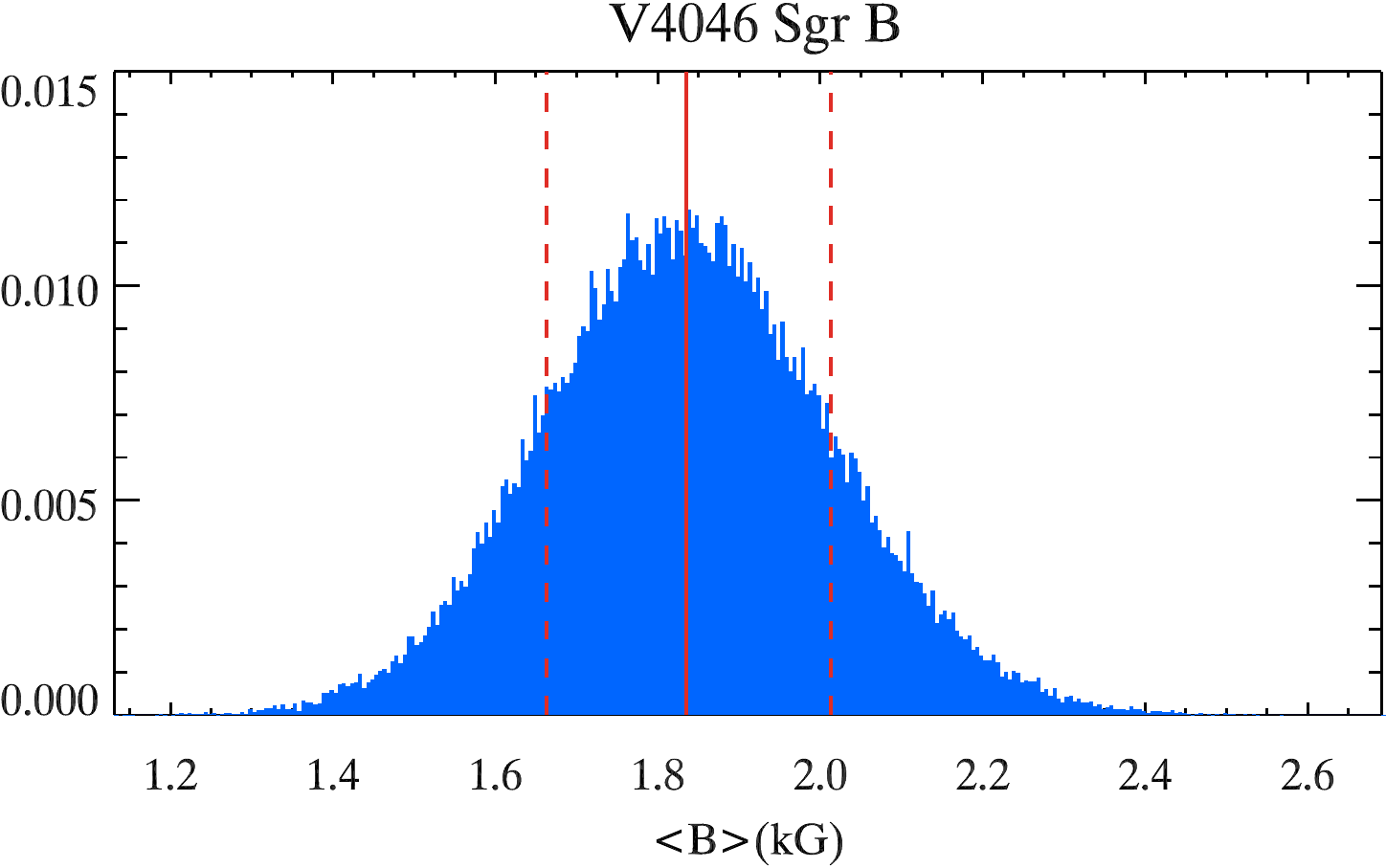}
\caption{Results for V4046 Sgr presented in the same way as for V1878 Ori in Fig.~\ref{fig:OriMagnetic}.}
\label{fig:SgrMagnetic}
\end{figure*}

The priors for V4046 Sgr were selected in a similar way as for V1878 Ori. The primary difference is that \cite{stempels:2004} estimated the luminosity ratio of the system to be $1.5$. While this value was determined in a different wavelength range than we used in the current study, it does indicate that there is a noticeable difference between the brightness of the two stars. Consequently, the uniform luminosity ratio prior was chosen as $1.5\pm0.5$. An abundance prior between $-6.90$ and $-7.80$ dex was selected. 
There seems to be a large radial velocity shift in the disentangled spectra of V4046 Sgr B. For this reason, its radial velocity was first adjusted by eye and was then allowed to vary in a range of $\pm2$~km\,s$^{-1}$. The prior for $v\sin{i}$ was set as $\pm2.5$~km\,s$^{-1}$ around the values used by previous studies \citep{quast:2000,stempels:2004}. 

Similarly to V1878 Ori, the most suitable model according to the BIC was a model containing three filling factors, corresponding to 0, 2, and 4~kG, for the magnetic field on each star (see Table \ref{tab:BIC}). The parameter values are reported in Table~\ref{tab:SgrResult}, with the best fit to observations and the average magnetic field posterior distributions shown in Fig. \ref{fig:SgrMagnetic}. Complete posterior distributions are also available in the appendix (Fig. \ref{fig:SgrResults}). \cla{The correlations between the inference parameters are qualitatively similar to those seen for V1878 Ori.} The mean magnetic fields at the surface of the two components are similar in strength: $1.96\pm0.18$ and $1.83\pm0.18$~kG for components A and B, respectively. 

The luminosity ratio obtained here is slightly lower than what was determined by \citet{stempels:2004}. This is likely explained by the fact that the observations analysed here correspond to a wavelength interval farther to the red compared to the analysis by \citet{stempels:2004}. The $v\sin{i}$ values that we derived are slightly higher than what has been obtained by previous studies \citep{stempels:2004, donati:2011}. The reason might be that other broadening parameters (such as macroturbulence) are not accounted for in our inference, which can lead to an overestimation of the stellar rotation rate.

\cla{Overall, our results for V4046 Sgr have larger uncertainties than for V1878 Ori. This difference comes from two sources. First, the S/Ns of the individual observations for V4046 Sgr are significantly lower than those for V1878 Sgr (see Table \ref{tab:OriSN} and \ref{tab:SgrSN}). The second reason is that the observations were then combined into the mean intensity spectra (see Sect. \ref{sec:SB}). There are in total 25 observations for V1878 Ori, but only eight for V4046 Sgr, which further contributes to the noise difference in the mean disentangled spectra.}

\begin{table}
\caption{Obtained parameters for the components of V4046 Sgr.}
\label{tab:SgrResult}
\begin{tabular}{lrr}
\hline \hline
 Parameter & \multicolumn{1}{c}{A} & \multicolumn{1}{c}{B} \\
\hline
    $\langle B_I \rangle$ (kG) & $1.96\pm0.18$ & $1.83\pm0.18$ \\
    $f_0$ & $0.42\pm0.07$ & $0.45\pm0.08$ \\
    $f_2$ & $0.17\pm0.12$ & $0.18\pm0.15$ \\
    $f_4$ & $0.40\pm0.07$ & $0.37\pm0.08$ \\
    $v\sin{i}$ (km\,s$^{-1}$) & $15.1\pm0.3$ & $14.4\pm0.3$\\
    $\varepsilon_{\mathrm{Ti}}$ & \multicolumn{2}{c}{$-7.68\pm0.03$} \\
    LR & \multicolumn{2}{c}{$1.27\pm0.04$} \\
    \hline
\end{tabular}
\tablefoot{$f_0$ is determined from Eq. \ref{eq:ff-constraint}.}
\end{table}

\subsection{Impact of microturbulent velocity}
\begin{table}
    \centering
    \caption{Inference results for V1878 Ori using different values of $v_\mathrm{mic}$.}
    \label{tab:vmic}
    \begin{tabular}{lrr|rr|rr}
        \hline \hline
        Parameter & \multicolumn{1}{c}{A} & \multicolumn{1}{c|}{B} & \multicolumn{1}{c}{A} & \multicolumn{1}{c|}{B} & \multicolumn{1}{c}{A} & \multicolumn{1}{c}{B}\\
        \hline
        $v_\mathrm{mic}$ (km\,s$^{-1}$) & \multicolumn{2}{c}{0.5} & \multicolumn{2}{|c|}{1.0} & \multicolumn{2}{c}{1.5} \\
        $\varepsilon_\mathrm{Ti}$ & \multicolumn{2}{c}{$-6.69$} & \multicolumn{2}{|c|}{$-6.93$} & \multicolumn{2}{c}{$-7.16$} \\
        $\ln{(Y|\hat{\theta})}$ & \multicolumn{2}{c}{$1507\pm2.7$} & \multicolumn{2}{|c|}{$1509\pm2.6$} & \multicolumn{2}{c}{$1505\pm2.6$} \\
        $\langle B_{I}\rangle$ (kG) & 1.25 & 1.40 & 1.33 & 1.57 & 1.67 & 1.94 \\
        \hline
    \end{tabular}
    
\end{table}

While our choice of $v_\mathrm{mic}$ is in line with the results of studies of other stars with similar atmospheric parameters, it would be interesting to assess how different choices would affect our results. To this end, we recalculated synthetic spectra for different values of $v_\mathrm{mic}$ in the range of 0.5--1.5 km\,s$^{-1}$ and repeated the magnetic inference described above. This analysis was only performed for the observations of V1878 Ori. Because the stellar atmospheres used for the two T Tauri systems are similar, the effect of changing $v_\mathrm{mic}$ should also be similar for the two binary systems. A few key parameters to evaluate the impact of different assumptions about $v_\mathrm{mic}$ are summarised in Table~\ref{tab:vmic}. 

Some variation between the outcomes of inferences with different $v_\mathrm{mic}$ was detected. Because the primary effect of $v_\mathrm{mic}$ is to increase the line depth, both the abundance and magnetic field strengths were affected. The Ti abundance varied by about 0.2 dex for a change in $v_\mathrm{mic}$ by 0.5\,km\,s$^{-1}$. For the same variation of $v_\mathrm{mic}$ , the magnetic field changed by 0.08--0.37\,kG, with a larger change corresponding to the variation of $v_\mathrm{mic}$ from 1.0 to 1.5~km\,s$^{-1}$. The quality of the fit with different $v_\mathrm{mic}$ values was also assessed using the likelihood (see parameter $\ln{(Y|\hat{\theta})}$ in Table~\ref{tab:vmic}). We found the likelihood to be highest for $v_\mathrm{mic}=1$~km\,s$^{-1}$. This difference in likelihood, however, is not significant when the 68\% confidence intervals of the inferences with different $v_\mathrm{mic}$ values are considered. To summarise, the set of \ion{Ti}{i} lines does not provide a strong independent constraint on $v_\mathrm{mic}$, likely because these lines do not differ much in terms of their intensities.

The uncertainty in the choice of $v_\mathrm{mic}$ increases the uncertainties of the inference results, particularly for the Ti abundance and the magnetic field strength. \cla{The additional uncertainty is larger than what was obtained from the inference (see Table~\ref{tab:OriResult}), which highlights the importance of accurate model assumptions. \cite{anderson:2010} also investigated the impact on magnetic field determination when changing stellar models. They considered the correlation of different stellar model parameters and average magnetic strength. In their results, they also found correlations between $v_\mathrm{mic}$ and $\langle B_{I}\rangle$, where a shift in a few hundred Gauss corresponded to a few tenths of km\,s$^{-1}$, similar to our results.}

However, this uncertainty does not have a significant impact on the relative difference between the magnetic field strengths of the two binary components. The relative field strength for V1878 Ori components, $\langle B_I\rangle_A/\langle B_I\rangle_B$, remains in the range of 0.85 to 0.9 for all three assumed $v_\mathrm{mic}$ values. For this reason, our analysis results should still be highly reliable at least for a comparison of the magnetic properties of the two components.

\section{Summary and discussion}
\label{sec:summary}

This work has presented the magnetic field analysis for two spectroscopic binaries, the T Tauri stars V1878 Ori and V4046 Sgr. By using multiple high-resolution observations, we were able to disentangle the spectra, remove telluric lines, and perform a detailed Zeeman-intensification analysis for each component. This allowed the determination of several stellar parameters as well as the properties of the small-scale magnetic field for the two components in the two binaries. By using the same observations as ZDI studies of these stars, we are now in the position to compare characteristics of the large- and small-scale magnetic fields more directly than was possible in previous studies \citep[e.g.][]{lavail:2019}.

Many previous studies showed that the magnetic field strengths recovered by ZDI are low, typically about 5-20\% compared to the field strength determined using Zeeman-intensification or -broadening analysis \citep{see:2019,kochukhov:2020a}. For T Tauri stars, the situation is slightly different; ZDI recovers more than 40\% of the magnetic field strength for some stars \citep{lavail:2019}. The mean magnetic field strengths for the components of V1878 Ori obtained using ZDI by \citet{lavail:2020} are 180 and 320~G for components A and B, respectively. Compared to our results of $1.33\pm0.08$ and $1.57\pm0.09$~kG, about 14 and 20\% of the magnetic flux from Zeeman intensification was recovered with ZDI for the respective components. For V4046 Sgr, the ZDI study by \citet{donati:2011} reported average global magnetic field strengths of 230 and 170~G for components A and B, respectively. Compared to our results of $1.96\pm0.18$ and $1.83\pm0.18$~kG, the ZDI magnetic fields correspond to about 12 and 9\% of the total magnetic flux revealed by our Zeeman-intensification analysis. These values are in line with previously established trends, although care should be taken when relations obtained for main-sequence stars are extended to pre-main-sequence objects. 

In the Zeeman-broadening study of T Tauri stars, \citet{lavail:2019} found that the field strength recovery capabilities of ZDI improve for simple, axisymmetric field geometries. The two components of V1878 Ori have relatively simple field geometries, and ZDI recovers a larger fraction of the total magnetic field than for the V4046 Sgr components, which have more complicated global field structures. However, \cite{lavail:2020} found that V1878 Ori A has a very weak contribution of axisymmetric magnetic field components (8.5\% of the total magnetic energy). This indicates that the simplicity of the magnetic field structure rather than degree of axisymmetry might be a more important factor in determining the ability of ZDI to infer a realistic total field strength. Because the spread between results obtained for similar stars is often quite large, more investigations about the field recovery fraction of ZDI in T Tauri stars could clarify these matters.

The average magnetic field strengths of the two components in the two binary stars are very similar. In addition, the magnetic filling factors corresponding to different magnetic field strengths are similar for the components (see Table \ref{tab:OriResult} and \ref{tab:SgrResult}). This further highlights the similarity between the properties of small-scale magnetic fields of the components because a similar filling factor distribution, in combination with a similar average field strength, indicates that the magnetic energies should be similar. On the other hand, the magnetic field strength distribution is different between V1878 Ori and V4046 Sgr. For the first system, the magnetic field is dominated by weaker field components, while the second system has higher contributions of stronger magnetic fields at the surface.

The similarity of the small-scale fields obtained for the binary components in this work can be compared with the significantly different global magnetic field strength and geometries obtained for V1878 Ori \citep{lavail:2020} and V4046 Sgr \citep{donati:2011} using ZDI. Lower-mass binary systems, with a mass ratio close to one, where both the global magnetic field geometry and the total field strength were investigated simultaneously \citep[e.g.][]{kochukhov:2017,kochukhov:2019}, also show smaller variations in the total strength obtained from the broadening and intensification of spectral lines than the average strength of the field recovered by ZDI. 
This pattern appears to be consistently present in binary stars of different sizes and evolutionary stages. Both partially and fully convective stars also seem to follow this trend. With similar initial conditions and mass, it is likely that the magnetic fields of the stars we studied are functionally similar, but that the constant evolution of the large-scale magnetic field geometry during magnetic cycles makes the ZDI result more dependent on when the star is observed rather than what type of star is observed. Repeated spectropolarimetric observations, such as those performed for single stars by \cite{boro-saikia:2018}, of binaries containing components with similar stellar parameters could clarify whether this dichotomy is just temporal in nature or a fundamental property of the different binary components. Combining the multi-epoch spectropolarimetric studies with simultaneous Zeeman-broadening or -intensification investigations could further improve the understanding of the magnetic field evolution of stars by determining how the overall magnetic energy changes over different timescales. 

In the near future, observations from the new generation of near-infrared spectropolarimeters such as SPIRou \citep{donati:2020} and CRIRES$^+$ \citep{dorn:2016} could give new insight into the magnetic fields of cool stars. With a high resolution and a large spectral grasp, these spectropolarimeters should be capable of providing data for simultaneous investigations of both small- and large-scale magnetic fields. The fact that these spectopolarimeters are operating at near-infrared wavelengths will also make any Zeeman-broadening signatures easier to detect. The result of this work highlights one question in particular that could be investigated by these future studies, that is, the similarity of the magnetic fields of stellar twins. Binary stars provide the best laboratories to answer this question as similar stellar parameters and origin reduce any potential variations that could otherwise impact the magnetic field generation, allowing a direct comparison.

\begin{acknowledgements}
\cla{We thank the referee for their comments.} This study is supported by the Swedish Research Council, the Swedish Royal Academy of Sciences, and the Swedish National Space Board.
The observational data analysed in this work were obtained at the Canada-France-Hawaii Telescope (CFHT), which is operated by the National Research Council of Canada, the Institut National des Sciences de l'Univers of the Centre National de la Recherche Scientifique of France, and the University of Hawaii.
\end{acknowledgements}

\bibliographystyle{aa}

\begin{appendix}
\label{appena}

\section{List of spectroscopic observations}
\begin{table}[h]
\caption{ESPaDOnS observations of V1878 Ori.}
\label{tab:OriSN}
\begin{tabular}{cc|cc}
\hline \hline
Reduced HJD & S/N & Reduced HJD & S/N\\
\hline
       57401.295& 180 & 57409.301& 176\\
       57401.462& 153 & 57410.380& 183\\
       57402.224& 181 & 57411.335& 191\\
       57402.451& 171 & 57412.361& 171\\
       57403.268& 188 & 57413.345& 182\\
       
       57403.449& 178 & 57414.383& 184\\
       57404.286& 188 & 57415.384& 181\\
       57404.451& 173 & 57416.344& 189\\
       57405.262& 196 & 57472.232& 153\\
       57405.378& 192 & 57474.232& 158\\
       
       57406.364& 192 & 57475.232& 143\\
       57407.287& 172 & 57476.232& 132\\
       57408.401& 159 & &\\
\hline
\end{tabular}
\end{table}
\newpage
\begin{table}[h]
\caption{ESPaDOnS observations of V4046 Sgr.}
\label{tab:SgrSN}
\begin{tabular}{cc}
\hline \hline
Reduced HJD & S/N \\
\hline
55077.273 & 66 \\
55078.316 & 64 \\
55079.221 & 75 \\
55080.220 & 78 \\
55080.313 & 80 \\
55081.221 & 60 \\
55082.222 & 78 \\
55083.221 & 82 \\
\hline

\end{tabular}
\end{table}
\newpage
\section{Inference results}
\begin{figure*}[b]
\centering
\includegraphics[width=\hsize]{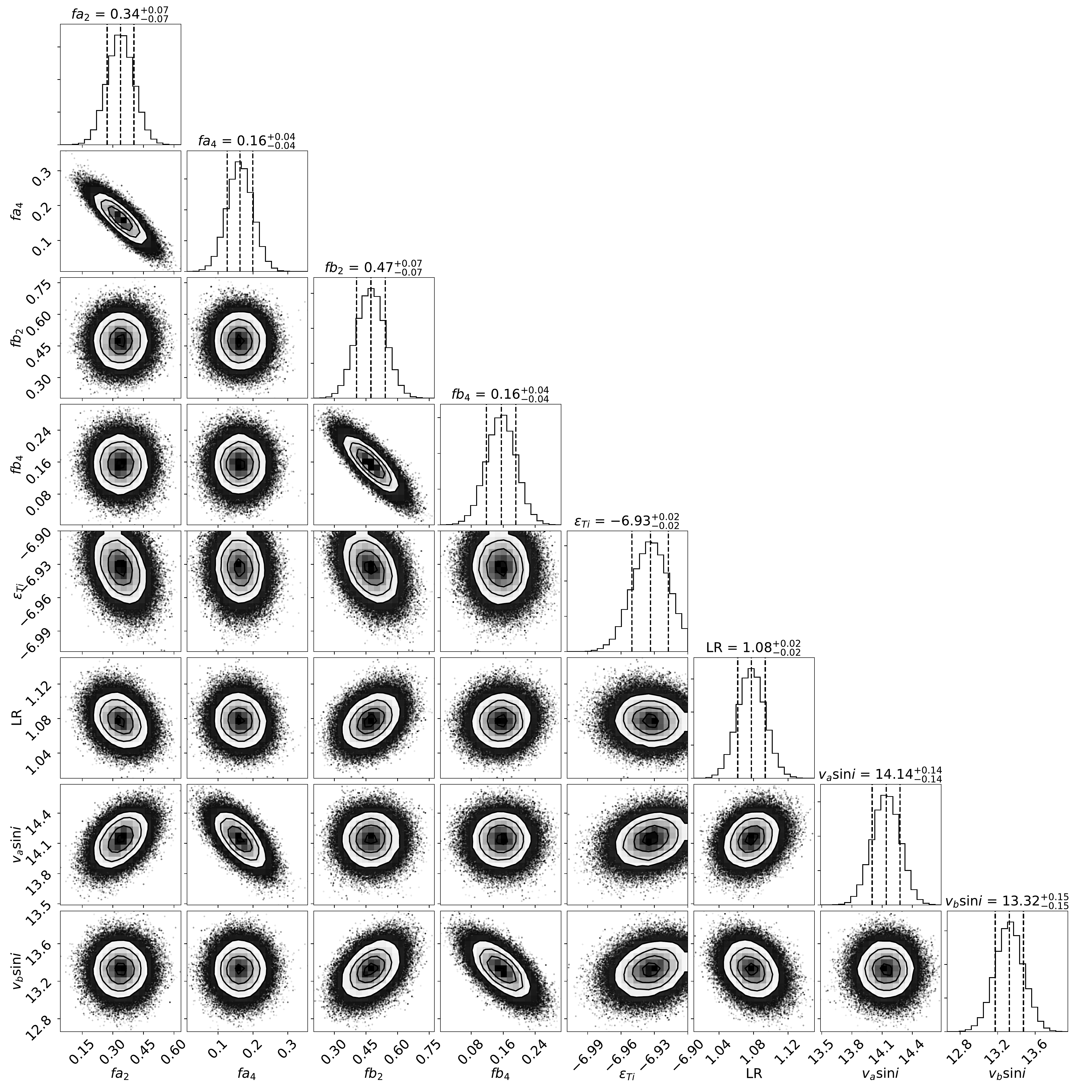}
\caption{Inference result for V1878 Ori showing the posterior distributions for the parameters derived with a simultaneous fit to the observed spectra of the two components.}
\label{fig:OriResults}
\end{figure*}

\begin{figure*}[b]
\centering
\includegraphics[width=\hsize]{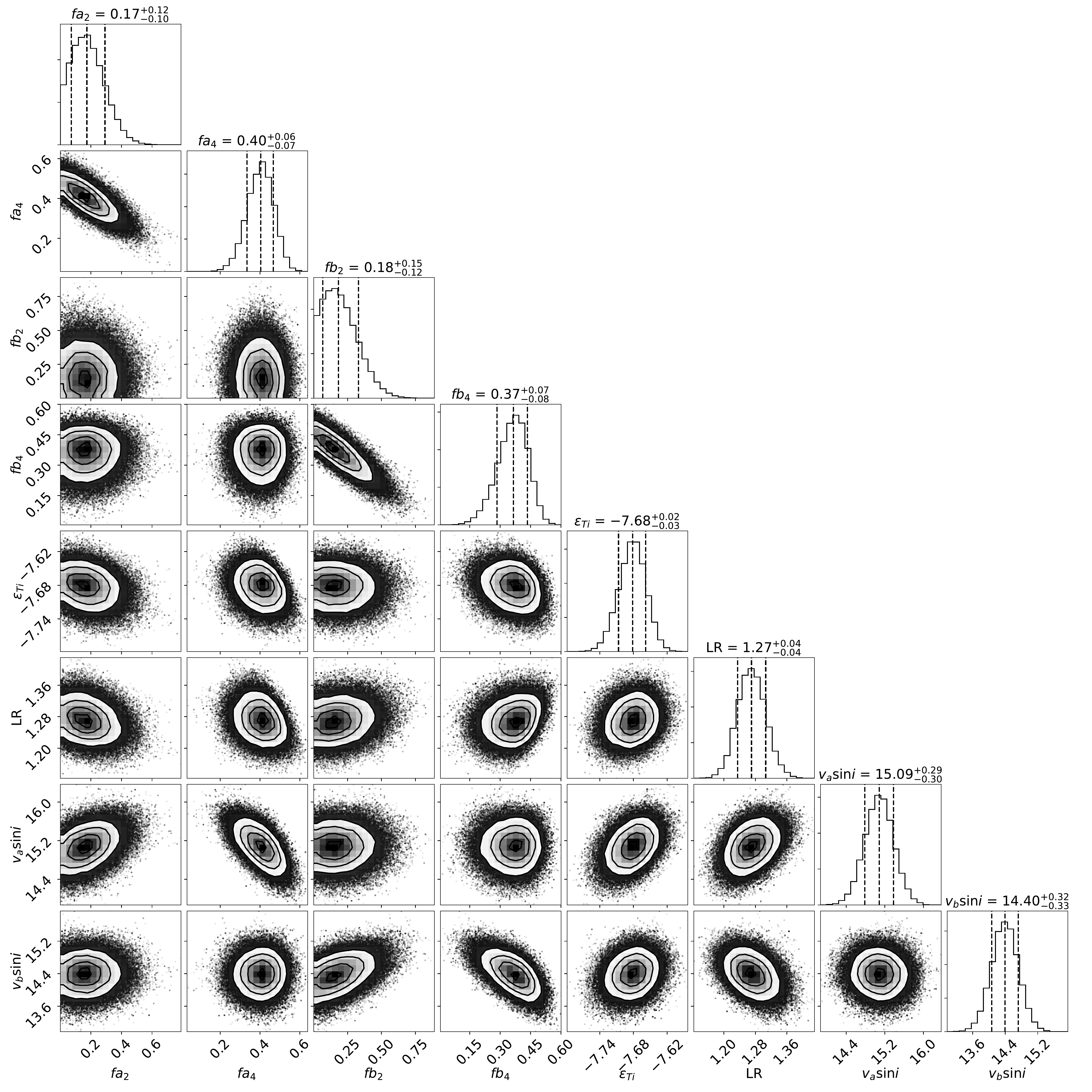}
\caption{Inference result for V4046 Sgr presented in the same way as for V1878 Ori in Fig.~\ref{fig:OriResults}.}
\label{fig:SgrResults}
\end{figure*}

\begin{table}[]
    \centering
    \caption{\cla{BIC values for different field parameterisation models.}}
    \begin{tabular}{ccc}
        \hline\hline
        $N$ & \multicolumn{2}{c}{BIC} \\
         & V1878 Ori & V4046 Sgr\\
        \hline
        2 & $-2109.60$ & $-1742.01$\\
        3 & $-2117.19$ & $-1743.16$\\
        4 & $-2070.95$ & $-1700.56$\\
        \hline
    \end{tabular}
    \label{tab:BIC}
\end{table}
\cla{Models containing up to four different magnetic field components were compared. This corresponds to a superposition of synthetic spectra computed for magnetic field strengths of 0, 2, 4, and 6\,kG. The resulting BIC values calculated using the Eq.~(\ref{eq:BIC}) are reported in Table \ref{tab:BIC}. The difference between the first two models are not particularly large, while the model with the most magnetic field components has noticeably higher BIC values. This indicates that including the 4\,kG component does improve the fit quality sufficiently to mitigate the complexity of the model, while adding the 6\,kG component has a very marginal effect on the fit quality. Figures \ref{fig:OriResults} and \ref{fig:SgrResults} show the posterior distributions of parameters for the model with the lowest BIC and three magnetic field components that were used to analyse the magnetic fields of the two binary stars.}

\end{appendix}
\end{document}